\documentclass[aps,prl,twocolumn,superscriptaddress,showpacs]{revtex4}
\usepackage{amssymb,graphicx,xspace,color}
\newcommand{\chris}[1]{\textcolor{black}{#1}}

\begin{document}

\title{Interband Quasiparticle Scattering in Superconducting LiFeAs Reconciles Photoemission and Tunneling Measurements} 

\author{Christian Hess}
\email[]{c.hess@ifw-dresden.de}
\affiliation{IFW Dresden, 01171 Dresden, Germany}
\author{Steffen Sykora}
\affiliation{IFW Dresden, 01171 Dresden, Germany}
\author{Torben H\"{a}nke}
\affiliation{IFW Dresden, 01171 Dresden, Germany}
\author{Ronny Schlegel}
\affiliation{IFW Dresden, 01171 Dresden, Germany}
\author{Danny Baumann}
\affiliation{IFW Dresden, 01171 Dresden, Germany}
\author{Volodymyr B. Zabolotnyy}
\affiliation{IFW Dresden, 01171 Dresden, Germany}
\author{Luminita Harnagea}
\affiliation{IFW Dresden, 01171 Dresden, Germany}
\author{Sabine Wurmehl}
\affiliation{IFW Dresden, 01171 Dresden, Germany}
\author{Jeroen van den Brink}
\affiliation{IFW Dresden, 01171 Dresden, Germany}
\affiliation{Department of Physics, TU Dresden, D-01069 Dresden, Germany}
\author{Bernd B\"{u}chner}
\affiliation{IFW Dresden, 01171 Dresden, Germany}
\affiliation{Department of Physics, TU Dresden, D-01069 Dresden, Germany}
\date{\today}

\begin{abstract}
Several angle resolved photoemission spectroscopy (ARPES) studies reveal a poorly nested Fermi surface of LiFeAs, far away from a spin density wave instability, and clear-cut superconducting gap anisotropies. On the other hand a very different, more nested Fermi surface and dissimilar gap anisotropies have been obtained from quasiparticle interference (QPI) data, which were interpreted as arising from \textit{intraband} scattering within hole-like bands. Here we show that this ARPES-QPI paradox is completely resolved by \textit{interband} scattering between the hole-like bands. The resolution follows from an excellent agreement between experimental quasiparticle scattering data and $T$-matrix QPI calculations (based on experimental band structure data), which allows disentangling interband and intraband scattering processes.

\end{abstract}


\maketitle 


The band structure of the iron-based superconductor LiFeAs became
recently a matter of strong controversial interest \cite{Allan2012,Borisenko2010,Umezawa2012,Borisenko2012,Kordyuk2011,Haenke2012,Putzke2012,Brydon2011,Platt2011,Ferber2012} since its precise knowledge is crucial for an understanding of the unusual superconductivity of LiFeAs. Iron-arsenides have the well-known property that experimentally observed factors such as Fermi surface (FS) nesting and van-Hove singularity-like behavior have a decisive impact on the type of spin fluctuations and on the superconducting ground state: A poorly nested model band structure of LiFeAs which mimics essential findings from angle resolved photoemission spectroscopy (ARPES) experiments yields prevailing ferromagnetic fluctuations with an instability towards triplet superconductivity \cite{Brydon2011}, whereas a calculated electronic band structure with a stronger nesting has been shown to result in an $s_\pm$-wave superconducting ground state driven by antiferromagnetic fluctuations \cite{Platt2011}.

Several ARPES studies on LiFeAs indeed consistently reveal a quite poor nesting, an (almost) van Hove singularity for one hole-like band at the Fermi level, and clear-cut momentum dependencies of the superconducting gap $\Delta_i({\bf k})$ for each band $i$ \cite{Borisenko2010,Umezawa2012,Borisenko2012,Kordyuk2011}. 
A much different normal state band structure which corresponds to a stronger nesting and other superconducting gap anisotropies $\Delta_i({\bf k})$ have, however, recently been deduced from a spectroscopic imaging scanning tunneling microscopy (SI-STM) study of the quasiparticle interference (QPI) of LiFeAs \cite{Allan2012}. This finding is based on the central assumption that quasiparticles inducing the QPI are scattered only internally within separate hole-like bands.

In this letter, we show that this seeming ARPES-QPI paradox is completely resolved when \textit{interband} scattering is taken into account. To be specific, we compare the experimentally observed QPI pattern at small wave vectors $|{\bf q}|\leq\pi/2$ with QPI calculations which  \textit{i)} are based on experimental band structure data and \textit{ii)} allow to consider individual scattering processes separately. We find an excellent agreement between the experimental and calculated data when the mentioned interband scattering is taken into account, whereas \textit{intraband} scattering alone (the essential hypothesis in Ref.~\onlinecite{Allan2012}) fails to yield such an agreement.
This explains the observed QPI's of Ref.~\onlinecite{Allan2012} and that of another QPI study of LiFeAs \cite{Haenke2012} without having to resort to low-energy band structure models that are contradicting ARPES experiments.

\begin{figure}[t]
\centering
\includegraphics[width=\columnwidth]{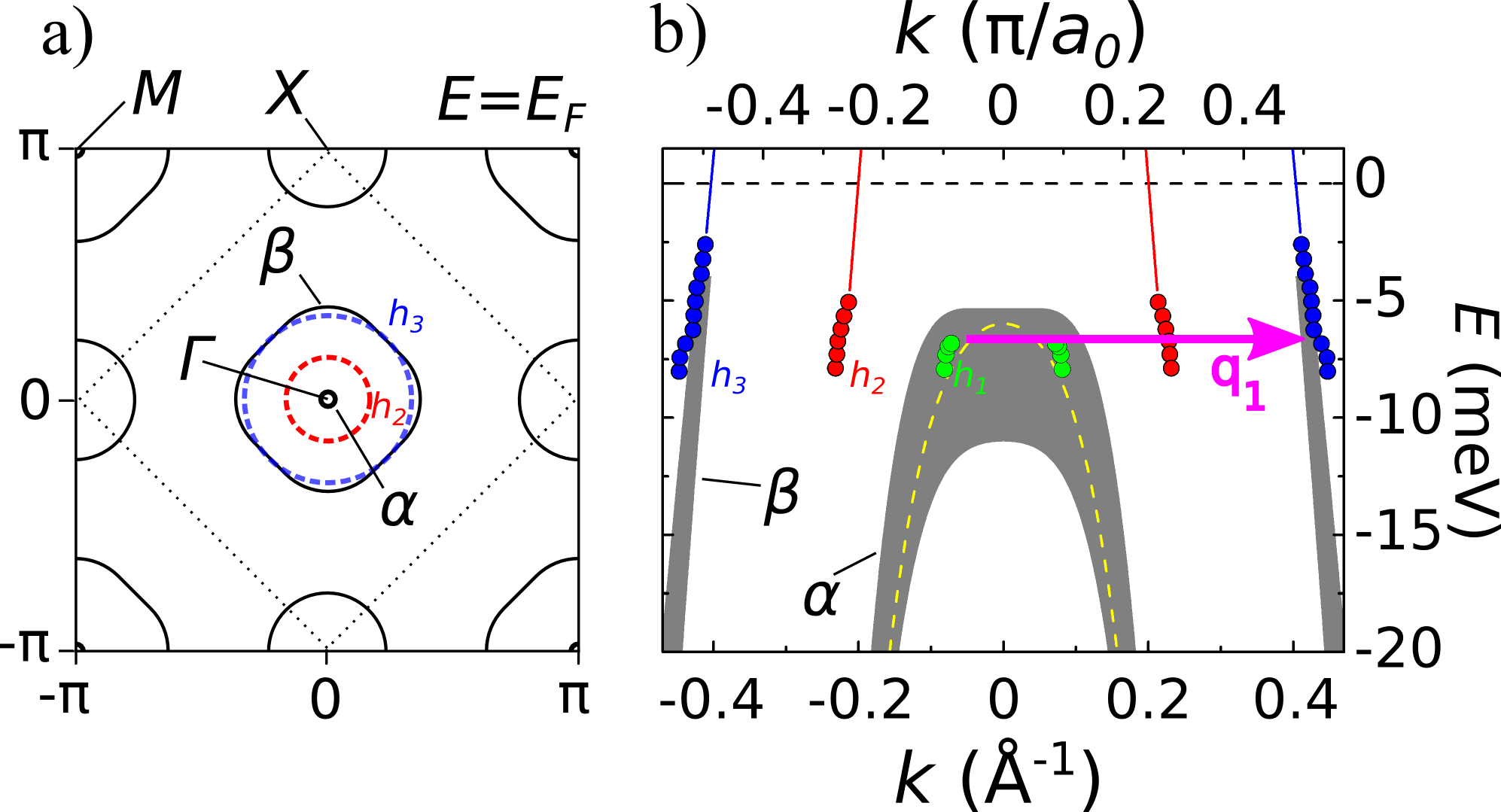}
\caption{a) Schematic illustration of the Fermi surface (FS) sheets of LiFeAs. Solid lines illustrate the approximate size of the FS sheets as observed by ARPES \cite{Borisenko2010,Kordyuk2011,Umezawa2012,Borisenko2012}, dashed lines correspond to the hole-like FS resulting from the hole-like bands $h_2$ and $h_3$ as obtained from QPI by Allan et al. \cite{Allan2012}. Note that Allan et al. extract the Fermi wave vectors along high-symmetry directions and that the shown FS sheets are respective isotropic two-dimensional extrapolations. The shown unfolded Brillouin zone (BZ) refers to the one-Fe unit cell. The BZ related to the two-Fe unit cell is indicated by dotted lines. b) LiFeAs band structure at negative energies. Circles represent the extracted quasiparticle dispersions $E_k$ of the hole-like bands $h_1$, $h_2$, $h_3$ obtained in Ref.~\onlinecite{Allan2012}, where $k$ is given by half of the length of the observed scattering vectors $\bf q$. Grey shaded contours represent $\alpha$ and $\beta$ bands as 
observed by ARPES \cite{Borisenko2010,Kordyuk2011,Umezawa2012,Borisenko2012}. The extended width of the $\alpha$ band  indicates $k_z$ dependence obtained from different photon energies \cite{Borisenko2012}. The indicated dispersion shown as a dashed line corresponds to data obtained at a photon energy $h\nu=20$~eV.
}
\label{FS_schem}
\end{figure}

LiFeAs is a \textit{stoichiometric} superconductor, i.e. superconductivity emerges already without doping, at a relatively high critical temperature $T_c\approx 18$~K \cite{Tapp2008}. 
Already this renders it radically different from the canonical '1111' and '122' iron-arsenide superconductors, where strong nesting between hole-like and electron-like FS pockets causes an antiferromagnetic spin density wave (SDW) ground state in the undoped parent compounds. There, superconductivity emerges only if the SDW state is weakened e.g. by doping \cite{Kamihara2006,Luetkens2009,Rotter2008b,Sefat2008a}, and nesting-related antiferromagnetic fluctuations have been suggested to drive the superconductivity with an  $s_\pm$-wave order parameter \cite{Mazin2008}.
LiFeAs, however, appears to be far away from an SDW instability: All doping attempts lead to a decrease of $T_c$ \cite{Aswartham2011a,Pitcher2010}, and
according to the ARPES studies \cite{Borisenko2010,Kordyuk2011,Borisenko2012,Umezawa2012}, the FS of LiFeAs consists of two similarly sized electron-like FS sheets around the $X$-point \footnote{Throughout this paper we refer to the one-Fe unit cell.}, and two hole-like FS sheets around the $\Gamma$-point arising from hole-like bands (labelled $\alpha$ and $\beta$, see Fig.~\ref{FS_schem}). One of the two bands has a relatively large FS ($\beta$-band, $k_F\approx0.4~\rm\AA^{-1}$), whereas the other is very small 
($k_F\lesssim0.1~\rm\AA^{-1}$) such that the corresponding $\alpha$-band just touches the Fermi level yielding a structure close to a van-Hove singularity. 
Due to the lack of surface states \cite{Lankau2010} these findings and those from SI-STM are expected to be bulk representative, and in fact the ARPES electronic structure has been shown to be well consistent with bulk sensitive Hall effect results \cite{Heyer2011}. 

In their QPI study \cite{Allan2012}, Allan et al. used the extracted scattering vectors of the QPI to construct three hole-band dispersions along high-symmetry directions, where they assumed the observed scattering vectors to stem from intraband scattering.
One of the resulting bands (labeled $h_3$ in Ref.~\onlinecite{Allan2012}) is consistent with the $\beta$-band observed in ARPES  \cite{Borisenko2010,Kordyuk2011,Borisenko2012,Umezawa2012}, and another ($h_1$) matches quite well the $\alpha$-band (see Fig.~\ref{FS_schem}). However, the third of the suggested bands ($h_2$) lacks such a correspondence since its Fermi wave vector $k_F\approx0.2~\rm\AA^{-1}$ \cite{Allan2012} neither matches that of the large nor the small hole-like FS observed in ARPES (see Fig.~\ref{FS_schem}). Its length is, however much closer to that of the electron-like bands and thus is the one which would result in a stronger nesting. 
 
\begin{figure}[t]
\centering           
\includegraphics[width=\columnwidth]{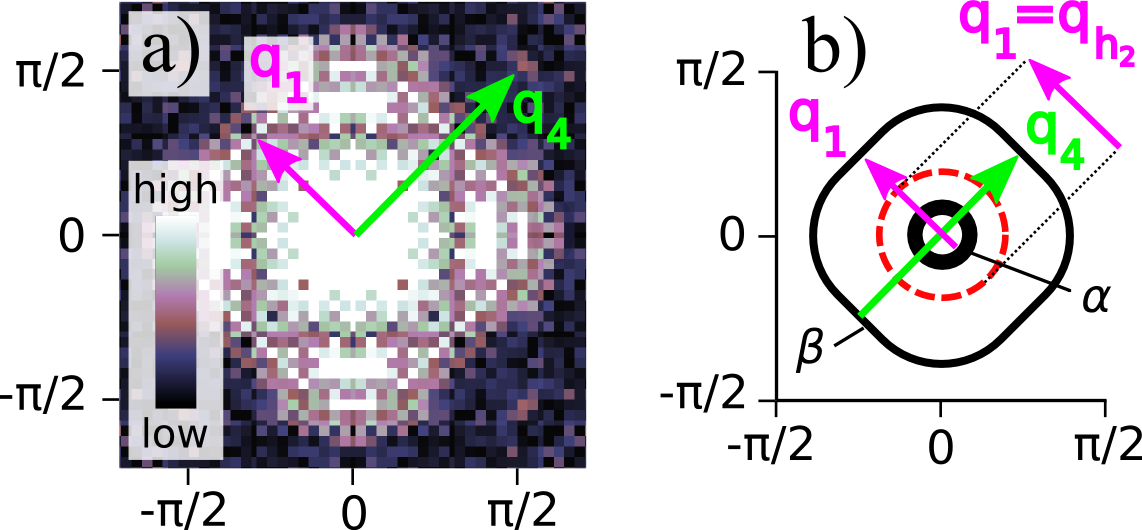}
\caption{a) Fourier transformed QPI data of LiFeAs at $E=-6.8$~meV \cite{Haenke2012}. b) Simplified constant energy contours (CEC) of the hole-like $\alpha$ and $\beta$ bands at $E=-6.8$~meV. ${\bf q}_1$ represents interband scattering processes which connect states on both bands. It has the same length as ${\bf q}_{h_2}$ (i.e., the diameter of the dashed $h_2$-CEC) reported in Ref.~\onlinecite{Allan2012}. ${\bf q}_4$ represents intraband scattering  within the $\beta$ band. 
\label{exp_schem}}
\end{figure}

As is evident in Refs.~\onlinecite{Allan2012} and \onlinecite{Haenke2012}, a prominent feature of the Fourier transformed experimental QPI pattern at negative bias is a bright squarish contour at small $|{\bf q}|\lesssim\pi/2$. 
Fig.~\ref{exp_schem}(a) shows our SI-STM data at $E=-6.8$~meV, which we have obtained on high-quality superconducting LiFeAs single crystals \cite{Morozov2010} in a home-built scanning tunneling microscope at $T=5.8$~K (see Ref.~\onlinecite{Haenke2012} for details of the measurement and the full topographic and spectroscopic data set). The chosen energy value is significantly larger ($>1$~meV) than the superconducting gap value \cite{Allan2012,Borisenko2012,Hanaguri2012,Umezawa2012}. Thus gap anisotropies do not play a strong role in the QPI pattern. The observed square structure can be straightforwardly understood by considering backscattering between the small ($\alpha$) and the large ($\beta$) hole band \cite{Haenke2012}: due to the high DOS of the $\alpha$-band and the relatively small momentum spread of this band at a given energy close to the band-top (see Fig.~\ref{FS_schem}(b)), the corresponding scattering vectors (labeled ${\bf q}_1$) just map the constant energy contour (CEC) of the $\beta$-band (
cf. Fig.~\ref{exp_schem}(b)). In order to underpin this statement, we show in Fig.~\ref{FS_schem}(b) that the diagonal ${\bf q}_1=(h,h)$ with $h=(0.285\pm0.05)\cdot{\pi/a}$ \cite{Haenke2012} with a length $|{\bf q}_1|=\sqrt{2}\cdot(0.285\pm0.05)\cdot\pi/a=(0.47\pm0.08)~\rm\AA^{-1}$ (with $a=a_0/\sqrt{2}=2.6809~\rm\AA$ the shortest Fe-Fe distance, and $a_0=3.7914~\rm\AA$ the in-plane lattice constant \cite{Tapp2008}) just connects states in the $\alpha$ and $\beta$ bands along the $\Gamma-M$ direction as observed at this energy by ARPES \cite{Borisenko2010,Borisenko2012,Kordyuk2011,Umezawa2012}.
Ref.~\onlinecite{Allan2012} reports a similar squarish structure, the size of which along the $\Gamma-M$ direction Allan et al. used to extract the $h_2$-dispersion. In fact, we find that the diagonal ${\bf q}_1$ perfectly matches the reported ${\bf q}_{h_2}\approx0.28\cdot2\pi/a_0=0.46~\rm\AA^{-1}$ at $E\approx -7~\rm meV$ \cite{Allan2012}. This shows unambiguously that \textit{i)} the present and the reported QPI patterns of Allan et al. are geometrically well consistent, and \textit{ii)} as a consequence, the suggested $h_2$-band by Allan et al. does not describe a quasiparticle dispersion but has to be reinterpreted as the dispersion of the scattering wave vectors ${\bf q}_1$ which connect the $\alpha$ and $\beta$ hole-like bands.

Having established this most important finding  we corroborate it 
further by QPI simulations which we use to analyze the influence of the 
individual possible scattering processes separately. The calculation of the
QPI intensity distribution arising from the local density of states (LDOS)
is based on a standard $T$-matrix approach. Thereby the underlying electronic
band structure of LiFeAs is modeled by a tight-binding approximation of 
the ARPES results of Refs.~\onlinecite{Borisenko2010} and \onlinecite{Borisenko2012}. 
From the latter work we have derived an appropriate model for
the superconducting gap function $\Delta_{\bf k} = 0.4~\mbox{meV} + 
6~\mbox{meV}\times\cos k_x \cos k_y$ which is able to model the measured
momentum dependence of the superconducting gap along the Fermi surface
of the hole pockets $\alpha$ and $\beta$ of Fig.~\ref{FS_schem}(a). The 
model describes anisotropic $s$-wave pairing where the gap value along
$\beta$ varies between $2.9$~meV and $3.9$~meV being minimal at the direction towards the 
electron-like FS. The gap along $\alpha$ is almost isotropic with 
a value of approximately $6$~meV.
\begin{figure}[t!]
\centering
\includegraphics[width=0.89\columnwidth]{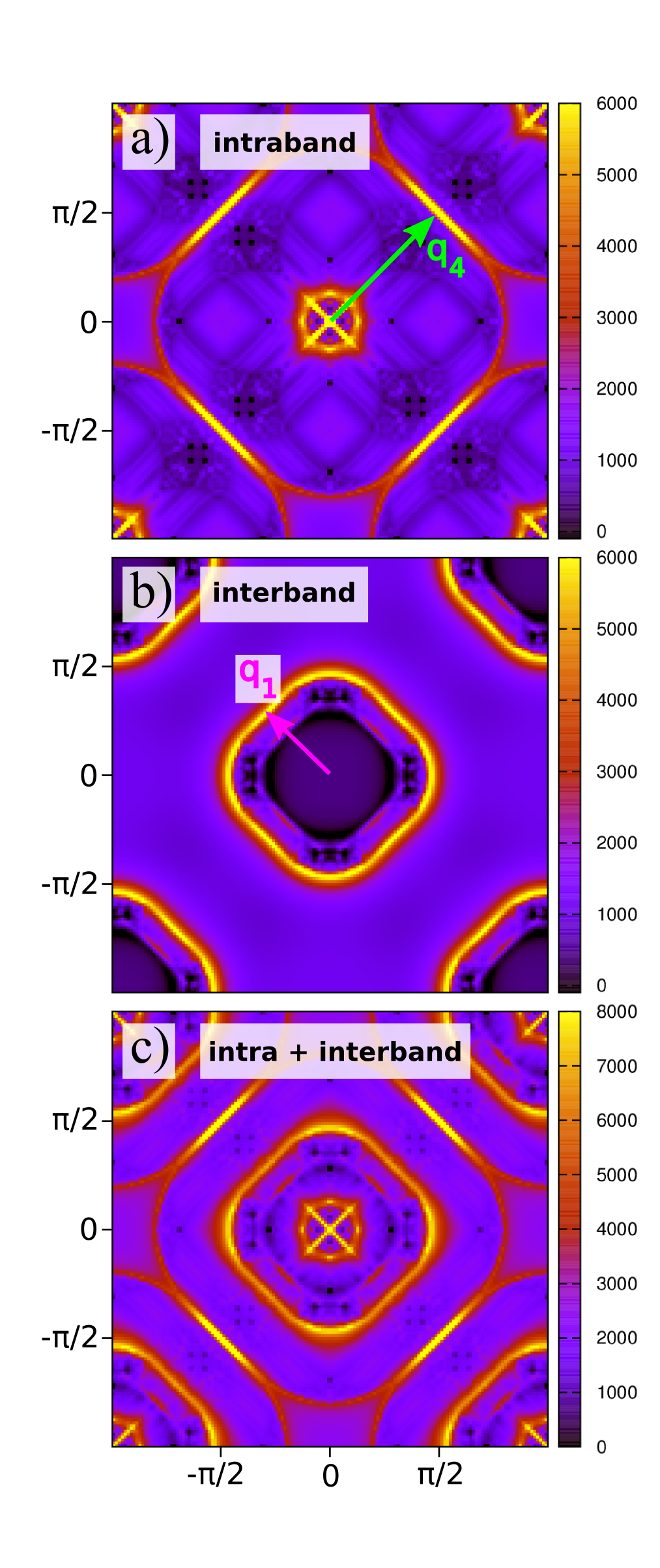}
\caption{Numerical simulation of the QPI patterns using a $T$-matrix approach
applied to a tight-binding model of the ARPES results 
\cite{Borisenko2010,Borisenko2012} and anisotropic $s$-wave pairing (see text).
a) Only intraband scattering within the small hole-like band
$\alpha$ and the large hole-like band $\beta$ is considered. All scattering
processes between $\alpha$ and $\beta$ and those processes involving the electron bands 
are suppressed in the calculation.  b) Only interband scattering between
$\alpha$ and $\beta$ is considered. c) Contributions displayed in panels a) and b) are
summed up in order to enable the comparison of the intensities. The scattering vectors ${\bf q}_1$, ${\bf q}_4$ (see Fig.~\ref{exp_schem}) are indicated.}
\label{theo}
\end{figure}

For the numerical evaluation we use a square lattice of $300 \times 300$ lattice 
points and periodic boundary conditions.  To simplify the treatment we assume that 
the isotropic scattering potential acts locally without 
coupling to the spin of conduction electrons (potential scattering). Furthermore, 
the cleanliness of the sample used for the STM measurements \cite{Haenke2012} allows
us to neglect multiple scattering processes (Born approximation). Fig.~\ref{theo} 
shows calculated QPI patterns for bias voltage $E=-6.8$~meV.
In order to identify 
the origin of the different features in the measured intensity distribution displayed in 
Fig.~\ref{exp_schem}(a) we faded out specific scattering processes such that
only intraband scattering within the two hole pockets $\alpha$ and $\beta$ (panel (a)) and 
only interband scattering between $\alpha$ and $\beta$ (panel (b)) entered the calculation
separately. In panel (c) the intensities of both scattering types are summed up.
As it is found already from geometrical interpretation of the experimental
result our numerical simulation confirms that the bright
squarish structure described by vector ${\bf q}_1$ has to be assigned clearly
to interband scattering. Moreover, its intensity is significantly larger than 
the intraband contributions which are seen as \textit{i)} a small region around the $\Gamma$
point indicating scattering within the small hole pocket $\alpha$ and \textit{ii)} a large
contour indicated by scattering vectors ${\bf q}_4$ 
connecting momenta within the large hole FS. The comparison between Fig.~\ref{theo}(c) 
and Fig.~\ref{exp_schem}(a) shows that the size and shape of most of the QPI 
features can be reproduced by our numerical simulation in excellent agreement 
with the experimental results. 

Beyond the bright structures we find also several
features showing minor intensity. For example, the intraband scattering within 
$\beta$ shows a pronounced maximum at momenta
$(\pm \pi/2,\pm \pi/2)$ along the lines connecting the points $(0,\pm \pi)$ 
and $(\pm \pi,0)$ which is seen very well in the experimental result (${\bf q}_4$). On the
other hand structures within the bright ${\bf q}_1$-contour appear with less intensity
as they show up in the experimental result Fig.~\ref{exp_schem}(a).
This difference might be traced back to the influence of coherence factors
entering the numerical simulation. It is well-known that a possible variation
of the superconducting phase along the Fermi surface strongly affects the
intensity distribution of the QPI at bias energy close to the superconducting
gap \cite{Haenke2012}. Note that within our simple anisotropic $s$-wave order 
parameter model though the gap value is consistent with ARPES studies
a possible variation of the phase is neglected completely. \chris{The consequence of phase variations would be QPI intensity which adds to the shown data, provided that the QPI is governed by potential scattering as is assumed in our calculation.}
Such phase effects
are interesting and will be subject to a forthcoming publication.

To sum up, we have shown that the quasiparticle interference in LiFeAs is dominated by interband scattering between two different hole-like bands around the $\Gamma$ point. This interpretation of the experimental results is supported by a comparison to a numerical simulation of the QPI based on a realistic tight-binding model that has been carefully extracted from recent ARPES data. Within this theoretical analysis we considered the 
different scattering processes separately and we obtained an excellent
agreement with the experimental QPI data from Refs.~\onlinecite{Haenke2012} and
\onlinecite{Allan2012}. This implies that a consistency between the results of the two experimental STM approaches and ARPES is obtained when going beyond considering only intraband scattering and include interband scattering explicitly. This changes fundamentally the interpretation of the extracted $h_2$-band and the proposed gap anisotropy  $\Delta_{h_2}({\bf k})$ of Ref.~\onlinecite{Allan2012}, and explains in both cases their strong discrepancy with ARPES data. For the $h_1$ and $h_3$ bands the consistency with the ARPES results and also the present QPI calculations is apparently very good. This is compelling evidence that the band structure suggested by APRES is indeed correct.
It is interesting to note, that this band structure is also well compatible with a recent de Haas van Alphen study \cite{Putzke2012}, where significant FS nesting in LiFeAs is claimed, despite only electron-like orbits in agreement with ARPES have been observed by this bulk sensitive technique so far.
\chris{We point out that even the incommensurate inelastic magnetic fluctuations which have recently been observed in volume sensitive inelastic neutron scattering experiments \cite{Qureshi2012} can very well be explained if inelastic scattering processes are simulated using our tight-binding band structure model \cite{Knolle2012}.
To conclude, the notion of LiFeAs being unique among the iron-arsenide superconductors is strongly reinforced by our study. Our main finding that the quasiparticle scattering in LiFeAs predominantly involves two separate hole-like Fermi surfaces may even have important consequences on the superconducting pairing mechanism and the tendency to form magnetically ordered states. According to Ref.~\onlinecite{Brydon2011}, small-momentum scattering vectors within the inner hole pocket combined with high density of states around the $\Gamma$ point would favour ferromagnetic fluctuations and a triplet pairing mechanism. Our study indeed confirms a relatively high intensity at small momenta in the QPI pattern. On the other hand, the highlighted interband scattering processes involving larger momentum vectors reveal a QPI structure of comparable high intensity. We note, however, that most of the scattering intensity still appears at relatively small momenta $|{\bf q}|<\pi/2$ which means that the argument of Ref.~\
onlinecite{Brydon2011} for triplet pairing is not modified by taking into account interband scattering.}

\section*{Acknowledgements}
The authors thank  S.V. Borisenko, J.C. Davis, I. Eremin, T. Hanaguri, D.K. Morr, A.W. Rost for valuable discussions and comments. Furthermore, we thank S.V. Borisenko for providing band structure data of LiFeAs. This work has been supported by the Deutsche Forschungsgemeinschaft through the Priority Programme SPP1458 (Grant No. BE1749/13 and GR3330/2) and the Graduate School GRK 1621.


\end{document}